
\magnification=\magstep1
\advance\voffset by 3truemm   
\advance\hoffset by -1truemm   
\vsize=23truecm  \hsize=16.5truecm
\overfullrule=0pt
\hfuzz 15truept
\parskip=5pt
\baselineskip=12pt
\font\tit=cmbx10 scaled \magstep2
\font\subt=cmbx10 scaled \magstep1
\font\ssubt=cmbx10
\font\auth=cmr10 scaled \magstep1
\font\titabs=cmti9
\font\abs=cmr9
\def\title#1{\null\vskip 26truemm \noindent {\tit #1} \vskip 12truemm
              \noindent {\auth By} }
\def\br{\hfil\break\noindent}  
\def\authors#1{\noindent {\auth #1}   \vskip 5truemm}
\def\address#1{\noindent #1\vskip 7truemm}
\def\abstract#1{\vskip 19truemm\noindent
   {\abs {\titabs Abstract} #1  }}
\def\section#1{\vskip 12truemm \noindent{\tit #1}\vskip 5mm\noindent}
\def\subsection#1{\vskip 8truemm \noindent{\subt #1}\vskip 3truemm\noindent}
\def\subsubsection#1{\vskip 5truemm \noindent{\ssubt #1}\vskip
2truemm\noindent}


%
%
%
 \def\redoffs{\voffset=-.31truein\hoffset=-.59truein}
\def\speclscape{\special{ps: landscape}}
%
%
%
%
\newbox\leftpage \newdimen\fullhsize \newdimen\hstitle \newdimen\hsbody
\tolerance=1000\hfuzz=2pt
\catcode`\@=11 
\def\bigans{b }
\message{ big or little (b/l)? }\read-1 to\answ
\ifx\answ\bigans\message{(This will come out unreduced.}
\else\message{(This will be reduced.} \let\l@r=L
\magnification=1000\baselineskip=16pt plus 2pt minus 1pt \vsize=7truein
\redoffs \hstitle=8truein\hsbody=4.75truein\fullhsize=10truein\hsize=\hsbody
\output={\ifnum\pageno=0 
  \shipout\vbox{\speclscape{\hsize\fullhsize\makeheadline}
    \hbox to \fullhsize{\hfill\pagebody\hfill}}\advancepageno
  \else
  \almostshipout{\leftline{\vbox{\pagebody\makefootline}}}\advancepageno
  \fi}
\def\almostshipout#1{\if L\l@r \count1=1 \message{[\the\count0.\the\count1]}
      \global\setbox\leftpage=#1 \global\let\l@r=R
 \else \count1=2
  \shipout\vbox{\speclscape{\hsize\fullhsize\makeheadline}
      \hbox to\fullhsize{\box\leftpage\hfil#1}}  \global\let\l@r=L\fi}
\fi
%
\newcount\yearltd\yearltd=\year\advance\yearltd by -1900

%
%

\def\draftmode{\message{ DRAFTMODE }\def\draftdate{{\rm preliminary draft:
\number\month/\number\day/\number\yearltd\ \ \hourmin}}%
\headline={\hfil\draftdate}\writelabels\baselineskip=20pt plus 2pt minus 2pt
 {\count255=\time\divide\count255 by 60 \xdef\hourmin{\number\count255}
  \multiply\count255 by-60\advance\count255 by\time
  \xdef\hourmin{\hourmin:\ifnum\count255<10 0\fi\the\count255}}}
\def\nolabels{\def\wrlabeL##1{}\def\eqlabeL##1{}\def\reflabeL##1{}}
\def\writelabels{\def\wrlabeL##1{\leavevmode\vadjust{\rlap{\smash%
{\line{{\escapechar=` \hfill\rlap{\sevenrm\hskip.03in\string##1}}}}}}}%
\def\eqlabeL##1{{\escapechar-1\rlap{\sevenrm\hskip.05in\string##1}}}%
\def\reflabeL##1{\noexpand\llap{\noexpand\sevenrm\string\string\string##1}}}
\nolabels
%
\global\newcount\secno \global\secno=0
\global\newcount\meqno \global\meqno=1
\def\newsec#1{\global\advance\secno by1\message{(\the\secno. #1)}
\global\subsecno=0\eqnres@t
\section{#1}}
\def\eqnres@t{\xdef\secsym{\the\secno.}\global\meqno=1\bigbreak\bigskip}
\def\sequentialequations{\def\eqnres@t{\bigbreak}}\xdef\secsym{}
\global\newcount\subsecno \global\subsecno=0
\def\subsec#1{\global\advance\subsecno by1\message{(\secsym\the\subsecno. #1)}
\ifnum\lastpenalty>9000\else\bigbreak\fi
\noindent{\it\secsym\the\subsecno. #1}\writetoca{\string\quad
{\secsym\the\subsecno.} {#1}}\par\nobreak\medskip\nobreak}
\def\appendix#1{\global\meqno=1\global\subsecno=0\xdef\secsym{\hbox{A.}}
\message{(#1)}
\section{#1}
\writetoca{\section{#1} }}
%
%
\def\eqnn#1{\xdef #1{(\secsym\the\meqno)}\writedef{#1\leftbracket#1}%
\global\advance\meqno by1\wrlabeL#1}
\def\eqna#1{\xdef #1##1{\hbox{$(\secsym\the\meqno##1)$}}
\writedef{#1\numbersign1\leftbracket#1{\numbersign1}}%
\global\advance\meqno by1\wrlabeL{#1$\{\}$}}
\def\eqn#1#2{\xdef #1{(\secsym\the\meqno)}\writedef{#1\leftbracket#1}%
\global\advance\meqno by1$$#2\eqno#1\eqlabeL#1$$}
%
\newskip\footskip\footskip14pt plus 1pt minus 1pt 
\def\footnotefont{\ninepoint}\def\f@t#1{\footnotefont #1\@foot}
\def\f@@t{\baselineskip\footskip\bgroup\footnotefont\aftergroup\@foot\let\next}
\setbox\strutbox=\hbox{\vrule height9.5pt depth4.5pt width0pt}
\global\newcount\ftno \global\ftno=0
\def\foot{\global\advance\ftno by1\footnote{$^{\the\ftno}$}}
%
\newwrite\ftfile
\def\footend{\def\foot{\global\advance\ftno by1\chardef\wfile=\ftfile
$^{\the\ftno}$\ifnum\ftno=1\immediate\openout\ftfile=foots.tmp\fi%
\immediate\write\ftfile{\noexpand\smallskip%
\noexpand\item{f\the\ftno:\ }\pctsign}\findarg}%
\def\footatend{\vfill\eject\immediate\closeout\ftfile{\parindent=20pt
\centerline{\bf Footnotes}\nobreak\bigskip\input foots.tmp }}}
\def\footatend{}
%
%
\global\newcount\refno \global\refno=1
\newwrite\rfile
\def\ref{[\the\refno]\nref}
\def\nref#1{\xdef#1{[\the\refno]}\writedef{#1\leftbracket#1}%
\ifnum\refno=1\immediate\openout\rfile=refs.tmp\fi
\global\advance\refno by1\chardef\wfile=\rfile\immediate
\write\rfile{\noexpand\item{#1\ }\reflabeL{#1\hskip.31in}\pctsign}\findarg}
\def\findarg#1#{\begingroup\obeylines\newlinechar=`\^^M\pass@rg}
{\obeylines\gdef\pass@rg#1{\writ@line\relax #1^^M\hbox{}^^M}%
\gdef\writ@line#1^^M{\expandafter\toks0\expandafter{\striprel@x #1}%
\edef\next{\the\toks0}\ifx\next\em@rk\let\next=\endgroup\else\ifx\next\empty%
\else\immediate\write\wfile{\the\toks0}\fi\let\next=\writ@line\fi\next\relax}}
\def\striprel@x#1{} \def\em@rk{\hbox{}}
\def\lref{\begingroup\obeylines\lr@f}
\def\lr@f#1#2{\gdef#1{\ref#1{#2}}\endgroup\unskip}

\def\addref#1{\immediate\write\rfile{\noexpand\item{}#1}} 
%

%
\def\startrefs#1{\immediate\openout\rfile=refs.tmp\refno=#1}
\def\xref{\expandafter\xr@f}\def\xr@f[#1]{#1}
\def\refs#1{\count255=1[\r@fs #1{\hbox{}}]}
\def\r@fs#1{\ifx\und@fined#1\message{reflabel \string#1 is undefined.}%
\nref#1{need to supply reference \string#1.}\fi%
\vphantom{\hphantom{#1}}\edef\next{#1}\ifx\next\em@rk\def\next{}%
\else\ifx\next#1\ifodd\count255\relax\xref#1\count255=0\fi%
\else#1\count255=1\fi\let\next=\r@fs\fi\next}
%

%
\newwrite\ffile\global\newcount\figno \global\figno=1
\def\fig{fig.~\the\figno\nfig}
\def\nfig#1{\xdef#1{fig.~\the\figno}%
\writedef{#1\leftbracket fig.\noexpand~\the\figno}%
\ifnum\figno=1\immediate\openout\ffile=figs.tmp\fi\chardef\wfile=\ffile%
\immediate\write\ffile{\noexpand\medskip\noexpand\item{Fig.\ \the\figno. }
\reflabeL{#1\hskip.55in}\pctsign}\global\advance\figno by1\findarg}
\def\xfig{\expandafter\xf@g}\def\xf@g fig.\penalty\@M\ {}
\def\figs#1{figs.~\f@gs #1{\hbox{}}}
\def\f@gs#1{\edef\next{#1}\ifx\next\em@rk\def\next{}\else
\ifx\next#1\xfig #1\else#1\fi\let\next=\f@gs\fi\next}
\newwrite\lfile
{\escapechar-1\xdef\pctsign{\string\%}\xdef\leftbracket{\string\{}
\xdef\rightbracket{\string\}}\xdef\numbersign{\string\#}}

\def\writestop{\def\writestoppt{\immediate\write\lfile{\string\pageno%
\the\pageno\string\startrefs\leftbracket\the\refno\rightbracket%
\string\def\string\secsym\leftbracket\secsym\rightbracket%
\string\secno\the\secno\string\meqno\the\meqno}\immediate\closeout\lfile}}
\def\writestoppt{}\def\writedef#1{}
\def\seclab#1{\xdef #1{\the\secno}\writedef{#1\leftbracket#1}\wrlabeL{#1=#1}}
\def\subseclab#1{\xdef #1{\secsym\the\subsecno}%
\writedef{#1\leftbracket#1}\wrlabeL{#1=#1}}
\newwrite\tfile \def\writetoca#1{}
\def\leaderfill{\leaders\hbox to 1em{\hss.\hss}\hfill}
\def\writetoc{\immediate\openout\tfile=toc.tmp
   \def\writetoca##1{{\edef\next{\write\tfile{\noindent ##1
   \string\leaderfill {\noexpand\number\pageno} \par}}\next}}}

%
\catcode`\@=12 
%
\edef\tfontsize{\ifx\answ\bigans scaled\magstep3\else scaled\magstep4\fi}
 \tfontsize  \tfontsize
 \tfontsize \font\titlei=cmmi10 \tfontsize
\font\titleis=cmmi7 \tfontsize \font\titleiss=cmmi5 \tfontsize
\font\titlesy=cmsy10 \tfontsize \font\titlesys=cmsy7 \tfontsize
\font\titlesyss=cmsy5 \tfontsize  \tfontsize
\skewchar\titlei='177 \skewchar\titleis='177 \skewchar\titleiss='177
\skewchar\titlesy='60 \skewchar\titlesys='60 \skewchar\titlesyss='60
 \ifx\answ\bigans\else scaled\magstep1\fi
\ifx\answ\bigans\else

 \font\absi=cmmi10 scaled\magstep1
\font\absis=cmmi7 scaled\magstep1 \font\absiss=cmmi5 scaled\magstep1
\font\abssy=cmsy10 scaled\magstep1 \font\abssys=cmsy7 scaled\magstep1
\font\abssyss=cmsy5 scaled\magstep1 
\skewchar\absi='177 \skewchar\absis='177 \skewchar\absiss='177
\skewchar\abssy='60 \skewchar\abssys='60 \skewchar\abssyss='60
\fi
\font\ninerm=cmr9 \font\sixrm=cmr6 \font\ninei=cmmi9 \font\sixi=cmmi6
\font\ninesy=cmsy9 \font\sixsy=cmsy6 \font\ninebf=cmbx9
\font\nineit=cmti9 \font\ninesl=cmsl9 \skewchar\ninei='177
\skewchar\sixi='177 \skewchar\ninesy='60 \skewchar\sixsy='60
\def\ninepoint{\def\rm{\fam0\ninerm}
\textfont0=\ninerm \scriptfont0=\sixrm \scriptscriptfont0=\fiverm
\textfont1=\ninei \scriptfont1=\sixi \scriptscriptfont1=\fivei
\textfont2=\ninesy \scriptfont2=\sixsy \scriptscriptfont2=\fivesy
\textfont\itfam=\ninei \def\it{\fam\itfam\nineit}\def\sl{\fam\slfam\ninesl}%
\textfont\bffam=\ninebf \def\bf{\fam\bffam\ninebf}\rm}
%
%

\hyphenation{anom-aly anom-alies coun-ter-term coun-ter-terms}
\def\inv{^{\raise.15ex\hbox{${\scriptscriptstyle -}$}\kern-.05em 1}}

\def\Dsl{\,\raise.15ex\hbox{/}\mkern-13.5mu D} 
\def\dsl{\raise.15ex\hbox{/}\kern-.57em\partial}

\def\lspace{\ifx\answ\bigans{}\else\qquad\fi}
\def\lbspace{\ifx\answ\bigans{}\else\hskip-.2in\fi} 
\def\boxeqn#1{\vcenter{\vbox{\hrule\hbox{\vrule\kern3pt\vbox{\kern3pt
	\hbox{${\displaystyle #1}$}\kern3pt}\kern3pt\vrule}\hrule}}}
\def\mbox#1#2{\vcenter{\hrule \hbox{\vrule height#2in
		\kern#1in \vrule} \hrule}}  
%

\def\darr#1{\raise1.5ex\hbox{$\leftrightarrow$}\mkern-16.5mu #1}

\def\roughly#1{\raise.3ex\hbox{$#1$\kern-.75em\lower1ex\hbox{$\sim$}}}

\title{GAUGE FIELD THEORIES ON RIEMANN SURFACES}
\authors{Franco Ferrari}\smallskip
\address{Dipartimento di Fisica, Universit\`a di Trento, 38050 Povo, Italy\br
and Istituto Nazionale di Fisica Nucleare, Gruppo Collegato di Trento, Italy}
\vskip 4cm
\abstract{
In this paper the free gauge field theories on a Riemann surface of any
genus are quantized in the covariant gauge.
The propagators of the gauge fields are explicitly derived
and their properties are analysed in details.
As an application, the correlation functions
of an Yang$-$Mills field theory
with gauge group $SU(N)$ are computed at the lowest order.}
\newsec{1 Introduction}

Recently, the quantized Yang$-$Mills field theories on Riemann surfaces
have been the subject of several investigations. A partial
list of the most relevant contributions is given in refs.
\ref\connections{M. Atiyah and R. Bott, {\it Phil. Trans. R. Soc. Lond.}
{\bf A 308} (1982), 523.}\nref\witten{
E. Witten,
{\it Comm. Math. Phys.}
{\bf 141}
(1991), 153;
{\it J. of Geom. and Phys.}
 {\bf 9}
(1992), 3781.
}\nref\gross{
D. J. Gross
{\it Nucl. Phys.}
{\bf B400}
(1993), 161; D. J. Gross and Washington Taylor IV,
{\it Nucl. Phys.}
{\bf B400}
(1993), 181;
{\it ibid.}
{\bf B403}
 (1993),
395.
}\nref\froe{J. Fr\"ohlich, On the Construction of Quantized
gauge Fields, in ``Field Theoretical Methods in Particle Physics" (W.
R\"uhl Ed.) Plenum, New York, 1980.}\nref\bt{
M. Blau and G. Thompson,
{\it Int. Jour. Mod. Phys.}
{\bf A7}
(1992), 3781; Lectures on the 2d Gauge Theories, lectures presented
at the 1993 Trieste Summer School in High Energy Physics, Preprint IC/93/356,
(hep-th/9310144);
G. Thompson,
1992 Trieste Lectures on Topological Gauge Theory and Yang$-$Mills Theory,
Preprint IC/93/356 and references therein.}\nref\hitchin{N. J.
Hitchin, {\it Gauge Theories on Riemann Surfaces},
in Lectures on Riemann Surfaces, M. Cornalba \& al. (eds), ICTP Trieste Italy,
9 Nov.-18 Dec. 1987, World Scientific editions; {\it Topology}, {\bf 31}
 (3) (1992), 449; {\it Proc. London
Math. Soc.} {\bf 55} (1987), 59.}\nref\rusa{B. Rusakov, {\it Mod.
Phys. Lett.} {\bf A5} (1990), 693.}--\ref\fine{
D. S. Fine,
{\it Comm. Math. Phys.}
{\bf 134}
(1990), 273; S. G. Rajeev, {\it Phys. Lett.} {\bf 212B}, 203; A. Sengupta,
{\it Ann. Phys.} {\bf 147} (1992), 191.}.
Despite of many important results, for instance the nonperturbative
computation of
the partition function and of the
amplitudes of the Wilson loops, the
possibility of performing explicit calculations in the case of
gauge fields interacting with matter is
confined until now
to the simplest topologies, like the cylinder, the disk,
the sphere and the torus
\ref\simptop{A. Z. Capri and R. Ferrari, {\it Nuovo Cimento} {\bf A 62} (1981),
273; {\it Jour. Math. Phys.} {\bf 25} (1983), 141;
N. Manton,
{\it Ann. Phys.}
{\bf 159}
(1985), 220; J. E.
Hetrick and Y. Hosotani,
{\it Phys. Rev.}
{\bf D38}
(1988), 2621;
A. K. Raina and G. Wanders,
{\it Ann. of Phys.} {\bf 132} (1981), 404;
C. Jayewardena, {\it Helv. Phys. Acta} {\bf 61} (1988), 636;
A. Bassetto and L. Griguolo, {\it Anomalous Dimensions and Ghost
Decoupling in a Perturbative Approach to the Generalized Chiral Schwinger
Model}, Preprint DFPD-94-TH-29, hep-th/9495953; A. Bassetto, F. de Biasio and
L. Griguolo, {\it Phys. Rev. Lett.} {\bf 72} (1994), 3141;
H. Joos, {\it Nucl. phys.} {\bf B17}
(Proc. Suppl.) (1990), 704; {\it Helv. Phys. Acta} {\bf 63} (1990),
670;
H. Dilger and H.
Joos,
How Well Do Lattice Simulations Reproduce the Different Aspects of the
Geometrical Schwinger Model, Contribution to the ``XI International
Symposium on Lattice Field Theory", Dallas 1993, Preprint DESY 93-144;
H. Joos and S. I.Azakov, The Geometric Schwinger Model on the Torus.2,
Preprint DESY-94-142;
K. S. Gupta, R. J. Henderson. S. G. Rajeev and O. T. Torgut,
{\it Jour. Math. Phys.} {\bf 35} (1994), 3845;
I. Sachs and A. Wipf, {\it Helv. Phys. Acta} {\bf 65} (1992),
653; E. Abdalla, M.C.B. Abdalla and K. D. Rothe, Nonperturbative
Methods in Two Dimensional Quantum Field Theory, World
Scientific, Singapore 1991.}.
On the other side,
the perturbative series of Yang$-$Mills theories can be derived
exploiting the powerful heat kernel techniques \ref\zfreg{B. S. DeWitt, The
Spacetime Approach to Quantum Field Theory, in: Relativity, Groups and
Topology II, B. S. DeWitt and R. Stora (eds), North Holland, Amsterdam,
pp. 381-738; G. Bernard, A. Duncan,
{\it Ann. Phys.} {\bf 107} (1977), 201; T. K. Leen, {\it Ann. Phys.}
{\bf 147} (1983), 417; I. L. Buchbinder and S. D. Odintsov, {\it Sov.
Phys.} {\bf J26} (1983), 359;
D. J. Toms, {\it Phys. Lett.} {\bf 126B} (1983), 37; {\it
Phys. Rev.} {\bf D27} (1983), 1803; L. Parker and D. J. Toms, {\it
Phys. Rev.} {\bf D29} (1984), 1584; M. D. Freeman, {\it Ann. Phys.}
{\bf 153} (1984), 339;  I. L. Buchbinder, S. D. Odintsov and I. L.
Shapiro, Effective Action in Quantum gravity, IOP Publishing, Bristol
1992; G. Cognola, L. Vanzo and S. Zerbini, {\it Phys. Lett} {\bf 241B}
(1990), 381;
G. Cognola, K. Kirsten and S. Zerbini, {\it Phys. Rev.} {\bf D48}
(1993), 790; E. Elizalde and S. D. Odintsov, {\it Phys. Lett.} {\bf
303B} (1993), 240; G. Cognola, {\it Phys. Rev.} {\bf D50} (1994),
909.}. For instance it is possible to check in this way
the renormalizability of any gauge field theory up to one loop approximation.
However, apart from the difficulty of performing calculations
at higher order, we are interested here in the explicit dependence of the
theory on the geometry of the Riemann surface, which is not so easy to treat
with heat kernel methods.
\smallskip
Consequently, in order to extend the investigations of refs. \simptop\
also to the case of Riemann surfaces,
we propose here a perturbative approach. One important
step in this direction is the construction of the propagator of the
gauge fields. To this purpose, we compute here the explicit expression
of the propagator in terms of theta functions and prime forms
\ref\fay{J. D. Fay, {\it Lect. Notes in Math.} {\bf 352}, Springer
Verlag, 1973.}.
Once the propagator is known, one can derive for instance the
vacuum expectation value (VEV) of the energy$-$momentum
tensor \ref\fulling{
N. D. Birrel and P. C. W. Davies, Quantum
Fields in Curved Space, Cambridge University Press, Cambridge (1982); S.
A. Fulling, Aspects of Quantum Field Theory in Curved Space-Time,
Cambridge University Press, Cambridge (1989).} at the lowest order.
In this calculation, the dependence on the moduli
of the two point function turns out to be crucial in
order to ascertain the existence of pseudoparticles in the physical
amplitudes.
The latter are
connected to the presence
of a gravitational background in certain local
systems of reference, see on this point refs. \ref\ferany{
F. Ferrari,
{\it Phys. Lett.} {\bf 277B} (1992), 423; {\it Comm. Math. Phys.} {\bf 156}
(1993), 179;
{\it
Int. Jour. Mod. Phys.} {\bf A9} (3) (1994), 313.}--\ref\fsu{F. Ferrari,
J. Sobczyk and W. Urbanik, Operator Formalism on the $Z_n$ Symmetric Algebraic
Curves, Preprint LMU-TPW 93-20, ITP UWr 856/93.}, where the example of free
conformal field theories is discussed.
The knowledge of the propagator alone, however, is not sufficient in
order to evaluate the radiative corrections of the correlation
functions on a Riemann surface because of the presence of zero modes
and of topologically nontrivial classical fields.
For this reason, we will give here
explicit formulas also for the flat connections following refs.
\bt, \hitchin\ and \ref\sonnen{J. Sonnenschein, {\it Phys. Rev.}
{\bf D42} (1990), 2080.}.
These connections play the role of external background fields,
so that the Yang$-$Mills theories on Riemann surfaces
can be treated within
the perturbative approach using the techniques
explained in refs.
\ref\background{L. F. Abbott, {\it Nucl. Phys.}
{\bf B185} (1981), 189; B. S. DeWitt, {\it Phys. Rev.} {\bf 162} (1967), 1195,
1239; G. 't Hooft, {\it Nucl. Phys.} {\bf B62} (1973), 444; H.
Kluberg-Stern and J. -B. Zuber, {\it Phys. Rev.} {\bf D12} (1975),
467, 482, 3159; S. D. Joglekar and B. W. Lee, {\it Ann. Phys. (NY)}
{\bf 97} (1976), 160.}.
As  a consequence, the final expression of the generating functional of the
one-particle irreducible Green functions will be gauge
invariant with respect to the background fields.
\smallskip
With the ingredients provided in this paper it is possible to start
the perturbative calculations of the $n-$point functions of many two
dimensional gauge field theories. Indeed,
even if the generating functional considered here involves
for simplicity only gauge fields, we are able to treat also
interactions with matter fields
without problems. Possible models are Yang$-$Mills field theories
interacting with massless fermions or scalars, for which the
propagators are already known from string theory.
Some of these theories are not exactly integrable, so that the use of
perturbative techniques is appropriate in these cases. On the other side,
if the theory is
integrable on the complex plane, nonperturbative
calculations can be achieved
also on Riemann surfaces once the free propagators are
known. An example concerning the Schwinger model
\ref\schwinger{J. Schwinger,
{\it Phys. Rev.} {\bf 128} (1962), 2425.}
has been given in
ref.
\ref\fersch{F. Ferrari, On the Schwinger Model
on Riemann Surfaces, Preprint LMU-TPW 93-26, MPI-Ph/93-71.}.\smallskip
In order to quantize the Yang$-$Mills field theory we choose here the
class of covariant gauge fixings $\nabla_\mu A^\mu=0$, where $\nabla_\mu$ is
the covariant derivative acting on the vector field $A^\mu$.
Unfortunately, due to the
presence of the metric, the equations of motion
satisfied by the Yang$-$Mills propagator are not easily solvable in this gauge.
A possible way out from this problem is to exploit the Lorentz gauge, in
which there is
the advantage that
the coexact components of the gauge fields decouple in the
free Lagrangian from the unphysical exact components. The linearized
equations of motion become equivalent to biharmonic equations
whose solutions is known on every Riemannian manifold
\ref\snwc{L. Sario,
M. Nakai, C. Wang and L. O. Chung, Lecture Notes in Mathematics {\bf 605},
Springer Verlag 1977.}. This is the strategy followed in ref.
\ref\ferabe{F. Ferrari, {\it Class. Q. Grav.} {\bf 10} (1993),
1065.} in the abelian case.
In the more complicated Yang$-$Mills field theories, however, the exact
components remain in the nonlinear part of the action, so that
the perturbative
expansion in the Lorenz gauge is very cumbersome.
For this reason we will use here another strategy, computing the
propagators after choosing on the Riemann surface a general,
but conformally flat metric. This is not a limitation, because every
metric on a
Riemann surface of given genus $h$ is conformally flat modulo global
diffeomorphisms.
Thus, the expressions given here for the propagators
can be extended to any other metric exploiting the invariance
under global
changes of coordinates of the Yang$-$Mills functional
quantized in the covariant gauge.\smallskip
Another problem to be solved in order to find the physical propagator
of the gauge fields
is that the Green functions obtained from the equations of
motion with a point source are not unique. The origin
of this nonuniqueness is the existence of the flat connections and the
residual gauge invariance typical of the covariant gauges.
The latter invariance can be related to the presence
of a constant zero mode in the free equations of motion \ferabe.
The arbitrariness
in the propagator is here removed imposing the condition
that the unphysical flat connections
should not be propagated inside the amplitudes. As we will see, this
requirement is sufficient also to eliminate the constant zero mode.
As a proof that our propagator is the physical one, we check that it
satisfies the Slavnov$-$Taylor identities
\ref\slatay{A. A. Slavnov, {\it Theor. Math. Phys.} {\bf 10} (1972), 99;
J. C. Taylor, {\it Nucl. Phys.} {\bf B33} (1971), 436.}
at the free level.
We notice that, on the contrary of what happens in string theory,
the 2-D Yang$-$Mills field theories are not conformally invariant.
Therefore, the only possible Slavnov-Taylor identities are those
associated to the gauge invariance of the theory.
\smallskip
The material contained in this paper is divided as follows.
In Section 2 we quantize the Yang$-$Mills field theories on a Riemann
surface in the covariant gauge using the BRST formalism
\ref\brst{C. Becchi, A. Rouet and R. Stora, {\it Comm.
Math. Phys.} {\bf 42} (1975), 127; {\it Ann. Phys.} {\bf 98} (1976),
287.}.
The equations defining the propagators of the gauge fields are
explicitly derived. They are too complicated to be solved for a
general metric, so that we limit ourselves to the conformally
flat metrics. We show however that the expression of the propagator can
be derived for any other metric exploiting the covariance of the
theory under general diffeomorphisms.
In Section 3 the two point functions of the ghost and gauge fields are
constructed. The already mentioned
arbitrariness given by the flat connections and by
the residual gauge invariance is totally eliminated by introducing three
physical requirements.
In Section 4 the properties of the gauge propagator
are investigated. First of all,
we verify that, on any open subset of the Riemann surface, it
is equivalent to the standard two point function of {\bf R}$^2$.
Secondly,  it is checked that the flat connections are not
propagated in the amplitudes. As a consequence of
the physicality of our propagator, we prove that its components fulfil
the Slavnov$-$Taylor identities
at the free level. Finally, for future applications in perturbation
theory, the structure of the divergent and finite parts
of the two point function at short distances is
computed. In Section 5 the generating functional of the correlation
functions for an $SU(N)$ Yang$-$Mills theory is considered.
The missing ingredient, the flat connections, are explicitly derived
in terms of the abelian differentials and
of the Lie algebra generators.
In Section 6 we present the conclusions and the possible future
developments.
Finally, the explicit  form of the components of the propagator
in the short distance limit is calculated in the appendix,
pointing out the differences
that appear considering Riemann surfaces of different genera.
\newsec{2 The Covariant Gauge Fixing on a Riemann Surface}

In this paper we consider the following Yang$-$Mills functional:
\eqn\sym{S_{YM}=\int_Md^2x\sqrt{g}\enskip {\rm Tr}\left[{1\over 4}F_{\mu\nu}
F^{\mu\nu}\right]}
where $M$ is a general closed and orientable Riemann surface of genus $h$
parametrized by the real coordinates $x_\mu$, $\mu=1,2$.
The metric on $M$ is given by the tensor $g_{\mu\nu}$ with Euclidean
signature and determinant denoted by $g\equiv{\rm det}[g_{\mu\nu}]$.
To fix the ideas, we suppose that the fields $A_\mu$ are elements of a
$su(N)$ algebra, so that $A_\mu\equiv \sum\limits_aA_\mu^aT^a$,
$a=1,\ldots,N^2-1$, where the $T^a$ are the generators of $SU(N)$ in
the adjoint representation.
The elements of the gauge group connected to the identity are
mappings $U(x):M\rightarrow SU(N)$ parametrized as follows: $U(x)={\rm exp}[
i\kappa\alpha^a(x)T^a]$. Here the $\alpha^a(x)$ represent real
functions on $M$ and $\kappa$ is a real coupling constant.
The field strength $F_{\mu\nu}$ appearing in eq. \sym\ is of the
form:
$$F_{\mu\nu}=\partial_\mu A_\nu-\partial_\nu A_\mu+i\kappa[A_\mu, A_\nu]$$
In this way it is easy to see that the action \sym\ is invariant under
a local $SU(N)$ transformation of the kind:
\eqn\gaugetransf{A_\mu(x)\rightarrow A_\mu^U(x)=
U^{-1}(x)\left[A_\mu(x)-i\kappa^{-1}\partial_\mu\right]U(x)}
To evaluate the trace in eq. \sym\ we will use the following conventions:
\eqn\trtatb{{\rm Tr}(T^aT^b)={1\over 2}\delta^{ab}}
\eqn\trtatbtc{{\rm Tr}(T^aT^bT^c)={1\over 4}(d^{abc}+if^{abc})}
where $d^{abc}$ is a totally symmetric tensor given by
$\{T^a,T^b\}=d^{abc}T^c$, with $\{,\}$ denoting the anticommutator, while
the $f^{abc}$ are the structure constants of the group $SU(N)$.
\smallskip
The classical action \sym\ is degenerate and in order to perform
the quantization we adopt the standard Faddeev and Popov procedure. To this
purpose, we introduce the set of nilpotent BRST
transformations \brst:
\eqn\brsone{\delta A_\mu^a=(D_\mu (A)c)^a\qquad\qquad\qquad\delta
B^a=0}
\eqn\brstwo{\delta c^a={1\over 2}f^{abc}c^bc^c\qquad\qquad\qquad
\delta\bar c^a={1\over 4}B^a}
where $\bar c^a$ and $c^a$ are the ghost fields and the $B^a$ play the
role of Lagrange multipliers.
The covariant derivative $D_\mu(A)$ appearing in eq. \brsone\ is
of the form: $D_\mu(A)=\nabla_\mu-i\kappa[A_\mu,\enskip]$.
Acting on the ghost scalar field $c(x)$, the differential operator
$\nabla_\mu$ is just the usual partial derivative
$\partial_\mu$.
After choosing a suitable gauge fixing $f^a(A)=0$, the total BRST
invariant action becomes:
\eqn\totalaction{S=S_{YM}+S_{GF}+S_{FP}}
where
\eqn\brsvar{S_{GF}+S_{FP}=\delta(\bar c^aF^a(A))}
is a pure BRST variation.
\smallskip
To our purposes, i.e. explicit perturbative calculations of the
correlation functions, it is convenient to impose the
covariant gauge fixing
\eqn\covgauge{
{1\over \sqrt{g}}
\partial_\mu(
\sqrt{g} A^\mu)=0
}
This preserves the covariance under general diffeomorphisms in the action
\totalaction. However,
the equations satisfied by the propagators of the gauge fields are
complicated by the presence of the
metric. An exception is provided by the Lorentz gauge already studied
in ref. \ferabe, where the coexact components of the $A_\mu$ fields
are completely decoupled from the exact components at the free level.
This fact allows the calculation of the propagator in a relatively
easy way and for any two dimensional manifold, also with boundary,
once
the biharmonic Green function with the proper boundary conditions
explained in \ferabe\ is known.
For this reason, the Lorentz gauge is very useful in treating some
models with abelian group of symmetry, like for instance the two dimensional
massless electrodynamics, in which the exact components can be simply
factored out from the path integral \fersch.
The situation is however different in the case of nonabelian gauge
field theories, because the exact components remain present in the
nonlinear interaction Lagrangian, making the perturbative approach in the
Lorentz gauge very cumbersome.
\smallskip
To solve the equations satisfied by the propagators in the covariant
gauge \covgauge\ we use the following strategy.
First of all, we introduce on $M$ a set of complex coordinates
$z=x_1+i x_2$, $\bar z=z^*$. Moreover, we exploit the fact that on a
Riemann surface it is always possible to choose a conformally flat
metric of the kind:
\eqn\confmetr{g_{zz}=g_{\bar z\bar z}=0\qquad\qquad\qquad g_{z\bar
z}=g_{\bar z z }=e^{\phi(z,\bar z)}}
$\phi(z,\bar z)$ being a real function.
At this point, we impose the gauge fixing \covgauge, which, in the particular
metric \confmetr, reduces to the simpler condition: $\nabla_z A_{\bar
z}+\nabla_{\bar z}A_z=0$.
This is a good gauge fixing apart from Gribov ambiguities
\ref\gribov{V. N. Gribov, {\it Nucl. Phys.} {\bf B139} (1978), 1.}, which we
will not discuss because our treatment is strictly perturbative.
Once the gauge invariance is fixed, the component $A^a_zdz$ ($A_{\bar
z}^ad\bar z$) of the gauge connection belongs to
$T^{*(1,0)}(M)$ ($T^{*(0,1)}(M)$), which is an holomorphic
(antiholomorphic) line bundle admitting holomorphic (antiholomorphic)
transition functions. As a consequence, the covariant gauge fixing \covgauge\
in a conformally flat metric reads:
\eqn\confgauge{f^a(A)=g^{z\bar z}(\partial_zA^a_{\bar
z}+\partial_{\bar z}A^a_z)=0}
Starting from the gauge fixed action \totalaction\ and integrating over
the Lagrange multipliers $B^a$ with the functional measure
$$d\mu[B]=\int DBe^{-{\lambda\over 32}\int_md^2x\sqrt{g}{\rm Tr}B^2}$$
we obtain the final formula:
\eqn\genfunct{Z[J]=\int DA_\mu D\bar cDc {\rm
exp}\left\{-{\rm Tr}\int_Md^2x\sqrt{g}
\left[{1\over 4}F_{\mu\nu}F^{\mu\nu}+{1\over
2\lambda}f^2(A)+\partial_\mu \bar cD^\mu(A)c+J_\mu A^\mu\right]\right\}}
{}From this generating functional it is possible to derive perturbatively
all the correlation
functions of the gauge fields in a conformally flat gravitational
background. This is not
a serious limitation, since the results can be easily extended to any other
metric in the following way.
For instance, let us suppose that the propagators are known
for a general metric $g_{z\bar z}$ which is conformally flat.
To derive the propagators
also in the case of another metric $\tilde g_{\mu\nu}(z',\bar z')$,
obtained
from $g_{z\bar z}$ after a global diffeomorphism plus a Teichm\"uller
deformation, it is sufficient to exploit the covariance under
diffeomorphisms of the action
appearing in eq. \genfunct.
This covariance is assured by the fact that the gauge fixing \confgauge\
is nothing but the covariant gauge \covgauge\ written in the
conformally flat metric \confmetr.
As an upshot, the classical equations of motion satisfied by the propagators
and the respective solutions turn out to be
covariant under global diffeomorphisms.
At this point, we notice that the metric
$\tilde g_{\mu\nu}(z',\bar z')$ is equivalent
to a conformally flat metric $g_{w\bar w}(w,\bar w)$ up to a change of
variables of the kind:
\eqn\chovar{w=w(z',\bar z')\qquad\qquad\qquad\bar w=\bar w(z',\bar z')}
(see for example \ref\naka{M. Nakahara, Geometry, Topology and Physics,
Adam Hilger, Bristol, Philadelphia and New York, 1990.} and references
therein).
In the new metric $g_{w\bar w}(w,\bar w)$ the components of the propagator
are known by hypothesis. Therefore, they can be computed also in the old
metric $\tilde g_{\mu\nu}(z',\bar z')$ inverting the diffeomorphism \chovar\
and using the covariance of the propagator mentioned above
under this transformation of
coordinates.
Concerning the other correlation functions, they are easily obtained
from the propagators exploiting perturbation theory.
Finally, let us notice that, in our perturbative framework, the addition
in eq. \genfunct\ of interactions with matter fields is not a problem.
For example, one may consider massless scalar or fermionic fields,
for which the propagators are already known from string theory.
\vfill\eject
\newsec{3 The Propagators in the Covariant Gauge}

Following the ideas of the previous section, we construct here the
propagators of the Yang$-$Mills field theory in the conformally flat
metrics described by eq. \confmetr.
To this purpose, it is sufficient to consider only the free part $S_0$
of the action appearing in eq. \genfunct. Using a set of complex
coordinates and dropping the color indices we obtain:
$$
S_0=\int_Md^2z
g^{z\bar z}
\left[
(\partial_z A_{\bar z})^2
\left (
{1\over \lambda}-1
\right)+
(\partial_{\bar z}
A_z)^2
\left(
{1\over \lambda}-1
\right)+
2\partial_zA_{\bar z}
\partial_{\bar z}
A_z\left({1\over \lambda}+1
\right)+\right.$$
\eqn\action{\left.
2J_zA_{\bar z}+2J_{\bar z}A_z\right]}
The classical equations of motion following from eq. \action\ are:
$$\left(1+{1\over \lambda}\right)\partial_{\bar z}(g^{z\bar z}\partial_z
A_{\bar z})-
\left(1-{1\over\lambda}\right)\partial_{\bar z}(g^{z\bar z}
\partial_{\bar z}A_z)
=J_{\bar z}$$
$$
\left(1+{1\over \lambda}\right)\partial_z(g^{z\bar z}\partial_{\bar z}
A_z)-
\left(1-{1\over\lambda}\right)\partial_z(g^{z\bar z}
\partial_zA_{\bar z})=J_z$$
As these equations show, the advantage of working with conformally
flat metrics is that the covariant derivatives are substituted by
partial derivatives, simplifying the calculations.
Now we are ready to compute the propagator of the gauge fields:
\eqn\propdef{G_{\alpha\beta}(z,w)\equiv
\langle A_\alpha(z,\bar z) A_\beta(w,\bar w)\rangle}
In eq. \propdef\ and in the following we adopt the conventions:
$\alpha=z,\bar z$, $\beta=w,\bar w$.
The equations defining the propagator become:
$$
\left(1+{1\over \lambda}\right)\partial_{\bar z}(g^{z\bar z}\partial_z
G_{\bar zw}(z,w))
-\left(1-{1\over\lambda}\right)\partial_{\bar z}(g^{z\bar z}
\partial_{\bar z}G_{zw}(z,w))=$$
\eqn\propone{\delta^{(2)}_{\bar zw}(z,w)-\sum\limits_{i,j=1}^h
\bar\omega_i(\bar z)\left[{\rm Im} \enskip
\Omega\right]^{-1}_{ij}\omega_j(w)}
$$\left(1+{1\over \lambda}\right)\partial_z(g^{z\bar z}\partial_{\bar z}
G_{z\bar w}(z,w))-
\left(1-{1\over\lambda}\right)\partial_z(g^{z\bar z}
\partial_zG_{\bar z\bar w}(z,w))=$$
\eqn\proptwo{\delta^{(2)}_{z\bar w}(z,w)
-\sum\limits_{i,j=1}^h
\omega_i(z)\left[{\rm Im} \enskip
\Omega\right]^{-1}_{ij}\bar\omega_j(\bar w)}
\eqn\propthree{\left(1-{1\over\lambda}\right)\partial_{\bar z}(g^{z\bar z}
\partial_{\bar z}G_{z\bar w}(z,w))-
\left(1+{1\over \lambda}\right)\partial_{\bar z}(g^{z\bar z}\partial_z
G_{\bar z\bar w}(z,w))=0}
\eqn\propfour{\left(1-{1\over\lambda}\right)\partial_z(g^{z\bar z}
\partial_zG_{\bar z w}(z,w))-
\left(1+{1\over \lambda}\right)\partial_z(g^{z\bar z}\partial_{\bar z}
G_{zw}(z,w))=0}
In the first two equations written above $\Omega_{ij}$, $i,j=1,\ldots,h$,
denotes the period
matrix and the $\omega_i(z)dz$ form a canonically normalized basis of
abelian differentials. Moreover, the
term in the right hand side of eqs. \propone\ and \proptwo\
is a projector onto the space of the zero modes, given in this case
by the $h$ abelian differentials $\omega_i(z)$, $i=1,\ldots,h$.
As shown in ref. \ferabe\
for the Lorentz gauge $\lambda=0$,
the presence of this projector is necessary
because otherwise also the
unphysical harmonic components of the fields would be propagated
in the amplitudes.
The proof that in the flat case
eqs. \propone-\propfour\ are equivalent
to the usual equations:
\eqn\flatdef{
\left[\delta_{\mu\nu}\triangle-\partial_\mu\partial_\nu\left(1-{1\over
\lambda}\right)\right]G_{\nu\rho}(x-y)=\delta_{\mu\rho}\delta^{(2)}(x-y)}
where $\triangle$ denotes the
Laplacian in cartesian coordinates and
$\mu,\nu=1,2$, is straightforward.\smallskip
At this point, we use eq. \propthree\ in order to derive the expression
of $G_{z\bar w}(z,w)$ in terms of $G_{\bar z\bar w}(z,w)$:
\eqn\gzbwdef{\partial_{\bar z}G_{ z\bar w}(z,w)={\left(\lambda+1\over
\lambda-1\right)}\partial_zG_{\bar z\bar w}(z,w)}
Substituting \gzbwdef\ in \proptwo, one
obtains an equation containing
only
$G_{\bar z\bar w}(z,w)$:
\eqn\gzbwbdef{
{\partial_z(g^{z\bar z}\partial_zG_{\bar z\bar w}(z,w))\over
(\lambda-1)}={1\over 4}\left[\delta^{(2)}_{z\bar w}(z,w)
-\sum\limits_{i,j=1}^h
\omega_i(z)\left[{\rm Im} \enskip
\Omega\right]^{-1}_{ij}\bar\omega_j(\bar w)\right]}
In the same way, from eqs. \propfour\ and \propone\ it follows that:
\eqn\gzwbdef{\partial_zG_{\bar z w}(z,w)={\left(\lambda+1\over
\lambda-1\right)}\partial_{\bar z}G_{zw}(z,w)}
and:
\eqn\gzwdef{
{\partial_{\bar z}(g^{z\bar z}\partial_{\bar z}G_{z w}(z,w))\over
(\lambda-1)}={1\over 4}\left[\delta^{(2)}_{\bar z w}(z,w)
-\sum\limits_{i,j=1}^h
\bar\omega_i(\bar z)\left[{\rm Im} \enskip
\Omega\right]^{-1}_{ij}\omega_j(w)\right]}
Thus we are left only with the two independent equations \gzbwbdef\ and
\gzwdef\ in the components $G_{zw}(z,w)$ and $G_{\bar z\bar w}(z,w)$.
To simplify these equations and to determine uniquely the form of the
propagator,
the following physical requirements play an important role:
\medskip
\item{a)} The unphysical zero modes should not be propagated.\medskip
\item{b)} The components of the propagator should be singlevalued.
Taking from example their integrals in $z$, the differential in $w,\bar w$
obtained in this way must not be periodic around the homology cycles $A_i$
and $B_i$, $i=1,\ldots,h$, of the Riemann surface:
\eqn\propertyb{\oint_\gamma dz G_{z\beta}(z,w)=\oint_\gamma d\bar z
G_{\bar z \beta}(z,w)=0}
where $\gamma$ is an arbitrary nontrivial homology cycle and $\beta=w,\bar
w$.
Analogous equations are valid integrating
in the variables $w,\bar w$.
\medskip
\item{c)} The gauge fields $A_z$ and $A_{\bar z}$ can be decomposed
according to the Hodge decomposition in coexact, exact and harmonic
components as follows:
\eqn\hodge{A_z=\partial_z\phi+\partial_z\rho+A^{\rm har}_z}
\eqn\hodgebar{A_{\bar z}=-\partial_{\bar z}\phi+
\partial_{\bar z}\rho+A^{\rm har}_{\bar z}}
Here $\phi$ is a purely complex scalar field, while $\rho$ is a real
scalar field. $A^{\rm har}_z$ and $A^{\rm har}_{\bar z}$
take into account the presence of the abelian differentials.
The decomposition \hodge\ and \hodgebar\ is not invertible unless
\eqn\consistency{\int_Md^2zg_{z\bar z}\phi(z,\bar z)=\int_Md^2zg_{z\bar
z}\rho(z,\bar z)=0}
Accordingly, also the propagator $G_{\alpha\beta}(z,w)$, which
from point a)
propagates only the coexact and exact components, should satisfy
analogous relations.\medskip\noindent
Applying the Hodge decomposition theorem \naka\ to the propagator \propdef,
one obtains that the most general form of this tensor is:
$$G_{\alpha\beta}(z,w)=\partial_\alpha\partial_\beta
G(z,w)+\sum_{i=1}^h\left(\partial_\alpha
f^iA^{iT}_\beta+A^{iT}_\alpha\partial_\beta f^{i\prime}+A_\alpha^{iT}
A^{iT}_\beta\right)$$
where $G(z,w)\equiv G(z,\bar z;w,\bar w)$,
$f^i(z,\bar z)$ and
$f^{i\prime}(w,\bar w)$ are
scalar functions and $A_\alpha^{iT}, A_\beta^{iT}$, $i=1,\ldots,h$,
represent a basis for the $2h$ real harmonic differentials
in the variables $(z,\bar
z)$ and $(w,\bar w)$ respectively. Let us notice that in the above
Hodge decomposition we ignored possible instantonic contributions.
In the Yang-Mills case they are ruled out by the fact that
the group $SU(N)$ is simply connected. In the abelian case, instead,
the instantonic gauge fields do not play any role because they
decouple in the free action $S_0$ from the exact and coexact
components.\smallskip
Clearly, requirements a) and b) are satisfied only if:
\eqn\gansatz{G_{\alpha\beta}(z,w)=\partial_\alpha\partial_\beta G(z,w)}
i.e. the harmonic components $A^{iT}_\alpha$ and $A^{iT}_\beta$
do not appear in the propagator.
The function $G(z,w)$ can be now computed exploiting the ansatz
\gansatz\ in eqs. \gzbwbdef\ and \gzwdef.
As an upshot, we obtain a biharmonic equation which
is solvable on any Riemann surface as shown
in ref. \ferabe\
for the particular case of the Lorentz gauge:
\eqn\bgf{G(z,w)=\int_Md^2t\sqrt{g}K(z,t)K(w,t)}
Here $K(z,t)$ is the well known scalar Green function defined by the
equations:
\eqn\kone{\partial_z\partial_{\bar z}K(z,t)=\delta^{(2)}_{z\bar z}(z,t)
-{g_{z\bar z}\over A}\qquad\qquad\qquad A=\int_Md^2zg_{z\bar z}}
\eqn\ktwo{\partial_{\bar z}\partial_tK(z,t)=-\delta^{(2)}_{\bar z
t}(z,t)+\sum\limits_{i,j=1}^h\bar \omega_i(\bar z)\left[{\rm
Im}\enskip\Omega\right]^{-1}_{ij}\omega_j(t)}
\eqn\kthree{\int_Md^2tg_{t\bar t}K(z,t)=0}
After a straightforward computation we obtain from eqs. \gzbwdef-\gzwdef\
the following final expressions for the components of the propagator:
\eqn\gzw{G_{zw}(z,w)=-{\lambda-1\over 4}\int_Md^2tg_{t\bar
t}\partial_zK(z,t)\partial_wK(w,t)}
\eqn\gzbw{G_{\bar zw}(z,w)=-{\lambda+1\over 4}\int_Md^2tg_{t\bar
t}\partial_{\bar z}K(z,t)\partial_wK(w,t)}
\eqn\gzbwb{G_{\bar z\bar w}(z,w)=-{\lambda-1\over 4}\int_Md^2tg_{t\bar
t}\partial_{\bar z}K(z,t)\partial_{\bar w}K(w,t)}
\eqn\gzwb{G_{z\bar w}(z,w)=-{\lambda+1\over 4}\int_Md^2tg_{t\bar
t}\partial_zK(z,t)\partial_{\bar w}K(w,t)}
It is now easy to check by direct substitution that the tensors
\gzw-\gzwb\ satisfy the equations of motion
\propone-\propfour\ identically.
In the proof, we have to permute the derivatives in $z$, $w$ or
in their complex complex conjugate variables with the
integrals in $d^2t$ appearing in eqs. \gzw-\gzwb.
This can be done without problems (see for example ref. \snwc)
because the components of the propagator given
above are well defined distributions.
As a matter of fact, they are
derivatives of the biharmonic Green function $G(z,w)$
which has been
extensively studied on any Riemannian manifold.\smallskip
We notice at this point that the
requirements a)-c), together with the free
equations of motion \propone-\propfour,
determine the propagator of the gauge fields
uniquely. Indeed, from a) and b) we obtained that the propagator should
be of the form \gansatz.
Moreover, from the equations of motion we were able to determine
the biharmonic Green function $G(z,w)$
up to solutions of the homogeneous biharmonic equation
$\triangle_g^2\varphi=0$ in $z$ and $w$, where $\triangle_g=2g^{z\bar
z}\partial_z\partial_{\bar z}$.
On a closed and orientable Riemann surface
this equation is equivalent to the following one:
\eqn\dfuc{\triangle_g\varphi={\rm constant}}
Now, it is well known that \dfuc\ does not admit any global
solution on $M$ apart from the trivial case in which the right hand
side vanishes  and $\varphi=\varphi_0$ is constant.
This possibility of adding a constant $\varphi_0$ to the biharmonic
Green function
is however ruled out by the conditions \consistency, which
require that the physical biharmonic Green function
satisfies the relations:
$$\int_Md^2g_{z\bar z}G(z,w)=\int d^2wg_{w\bar w}G(z,w)=0$$
It is in fact easy to see with the help of \kthree\
that the function $G(z,w)+\varphi_0$ verifies
the above equations only if $\varphi_0=0$.
\smallskip
Before concluding this section, two remarks are in order.
First of all, we notice that eqs. \gzw-\gzwb\ yield the explicit
form of the components of the gauge propagators on a Riemann surface
of any genus for the class of covariant gauges \covgauge.
As a matter of fact,
the expression of $K(z,t)$ in terms of the prime form and of
the abelian differentials is known on every closed and orientable
Riemann surface \ref\vv{E. Verlinde and H. Verlinde, {\it Nucl. Phys.}
{\bf B288} (1987), 357.} and can be explicitly constructed also
on algebraic
curves \fay,
\ref\ferstr{F. Ferrari, {\it Int. Jour. Mod. Phys.} {\bf A5}
(1990), 2799.}.
Moreover the propagator \propdef\ computed here is a well defined
tensor on $M$.
Exploiting its covariance under diffeomorphisms
in the two indices $\alpha$ and $\beta$ it is possible to extend the
calculations performed here also to a general metric as explained in
the previous section.\smallskip
To complete our discussion, we have to derive the propagator
$G_{gh}(z,w)$ of the ghost fields. From eq. \genfunct\ it turns out
that this Green function satisfies at the lowest order the following
harmonic  equation:
\eqn\eqmotgh{\triangle_g G_{gh}(z,w)=\delta^{(2)}(z,w)-{g_{z\bar z}\over
A}}
The term $1/A$ is required by the presence of a constant zero mode.
Comparing with eq. \kone, it is clear that
\eqn\ggh{G_{gh}(z,w)=K(z,w)}
\newsec{4 Further Properties of the Propagator}

First of all we verify that, locally, the components of the
propagator computed in the previous section coincide with the flat
ones.
To prove this fact we start with the well known flat
propagator, written
in real coordinates $x=(x_1,x_2)$ and $y=(y_1,y_2)$:
\eqn\flatprop{G_{\mu\nu}(x,y)={\delta_{\mu\nu}\over
\triangle}+(\lambda-1){\partial_\mu\partial_\nu\over \triangle^2}}
Formally, this propagator satisfies eq. \flatdef.
We compute now the components of \flatprop\ in complex coordinates exploiting
the conventions:
\eqn\cmplxcomp{\cases{z=x_1+ix_2\cr \bar z=x_1-ix_2\cr}\qquad\qquad\qquad
\cases{\partial_z={1\over 2}(\partial_1-i\partial_2)\cr
\partial_{\bar z}={1\over 2}(\partial_1+i\partial_2)\cr}}
After a few calculations one finds:
\eqn\flatgzw{G_{zw}(z,w)={1\over
4}\left[G_{11}-G_{22}-i(G_{12}+G_{21})\right](x,y)=-(\lambda-1){\partial_z
\partial_w\over
\triangle^2}}
and, analogously:
\eqn\flatgzwb{G_{z\bar w}(z,w)=-(\lambda+1){\partial_z\partial_{\bar w}
\over \triangle^2}}
\eqn\flatzbw{G_{\bar zw}(z,w)=-(\lambda+1){\partial_{\bar z}\partial_w
\over \triangle^2}}
\eqn\flatgzbwb{G_{\bar z\bar w}(z,w)=-
(\lambda-1){\partial_{\bar z}\partial_{\bar w}\over \triangle^2}}
In deriving the above equations we have used the translational invariance
of the flat Green functions,  so that $\partial_z{1\over
\triangle}=-\partial_w{1\over \triangle}$
and so on for the derivatives $\partial_{\bar z}$
and $\partial_{\bar w}$.\smallskip
On the other side, the scalar Green
function $K(z,t)$ appearing in eq. \kone\ is proportional to the inverse
of the Laplacian $\triangle_g$ defined on the Riemann surface,
i.e. $K(z,t)\equiv {2\over \triangle_g}$.
This can be seen from eq. \kone\ noting that,
in complex coordinates, $g^{z\bar
z}\partial_z\partial_{\bar z}={\triangle_g\over 2}$.
Thus it follows that the biharmonic Green function $G(z,w)$ introduced in
eq. \bgf\ is equal to $4\over \triangle_g^2$.
At this point it is easy to check that the components \flatgzw-\flatgzbwb\
obtained from the flat propagator \flatprop\ are equivalent to those of eqs.
\gzw-\gzwb\ on any open patch $U$ of $M$. For example, from eq. \gzw\
it is possible to rewrite $G_{zw}(z,w)$ in the following way:
$$G_{zw}(z,w)=-(\lambda-1){\partial_z\partial_w\over \triangle_g^2}$$
Choosing on $U$ a locally flat metric, we have $\triangle_g=\triangle$
and the above equation coincides with
\flatgzw. Analogous identities arise in the case of the remaining
components completing our proof.
\smallskip
Next, we verify the compatibility of the propagator derived in
Section 2 with requirements a)-c).
The proof of a) is very simple. The components
of the  propagator are in fact exact or coexact differentials in $z$ and $w$,
so that one can exploit the orthogonality properties of
the Hodge decomposition stating that the exact and coexact
differentials are always orthogonal with respect
to the abelian differentials on  $M$ \naka.
Therefore, using the standard definition of the scalar
product between one$-$forms, one obtains:
$$\int_Md^2zG_{z\beta}(z,w)\bar \omega_i(\bar z)=
\int_MG_{\bar z\beta}(z,w) \omega_i(z)=0$$
for $i=1,\ldots,h$ and $\beta=w,\bar w$.
Analogous equations are valid in the variables $w$ and $\bar w$ proving
requirement a).
Also the singlevaluedness of the propagator, in particular eq.
\propertyb\ of point b), is a direct consequence of the
form of the components \gzw-\gzwb, which are total derivatives
of the biharmonic function \bgf\ with respect to the variables
$z,w$ and their complex conjugates.
Finally, eq. \consistency\ follows from eq. \kthree\ as already shown
in the previous section.
\smallskip
Since the propagator is uniquely fixed by the equations of motion and
by the physical requirements a)-c), it should also satisfy the Slavnov-Taylor
identities associated to the BRST invariance of the gauge fixed theory
\genfunct\ under the transformations \brsone\ and \brstwo.
In particular, let us consider the Green function $\langle A_\alpha^a(z,\bar z)
\bar c^b(w,\bar w)\rangle$:
$$0=\delta\langle A_\alpha^a(z,\bar z)
\bar c^b(w,\bar w)\rangle=\langle\left(\partial_\alpha c^a(z,\bar z)-\kappa
f^{ade}c^d(z,\bar z)A^e_\alpha(z,\bar z)\right)\bar c^b(w,\bar w)
\rangle$$
$$-{1\over
\lambda}\langle A_\alpha^a(z,\bar z)\partial_\beta A^{\beta b}(w,\bar
w)\rangle$$
Applying the operator $\partial^\alpha$ to both sides of the above
equation  and keeping only the zeroth order terms with respect to the coupling
constant $\kappa$, we obtain the identity:
\eqn\fbrseq{{1\over \lambda}\partial^\alpha\partial^\beta
G_{\alpha\beta}(z,w)=-\delta^{(2)}(z,w)+{g_{z\bar z}\over A}}
The right hand side has been computed exploiting the equations of
motion of the ghost fields \eqmotgh.
At this point we substitute in eq. \fbrseq\ the components of the
propagator \gzw-\gzwb\ derived before.
Eqs. \gzbwdef\ and \gzwbdef\ yield:
\eqn\intone{\partial^\alpha\partial^\beta
G_{\alpha\beta}(z,w)=g^{z\bar z}g^{w\bar w}{2\lambda\over
\lambda-1}\left[\partial_z\partial_w G_{\bar z\bar w}(z,w)+\partial_{\bar
z}\partial_{\bar w}G_{zw}(z,w)\right]}
Using the fact that
$$\partial_wG_{\bar z\bar w}(z,w)=-{(\lambda-1)\over 4}g_{w\bar
w}\partial_{\bar z}K(z,w)$$
$$\partial_{\bar w}G_{zw}(z,w)=-{(\lambda-1)\over 4}g_{w\bar
w}\partial_{z}K(z,w)$$
and with the help of eq. \kone, it is easy to see that \intone\ is
nothing but the Slavnov-Taylor identity \fbrseq.\smallskip
To conclude this section, we
compute the structure of the singularities in the components
of the propagator. In view of perturbative applications, in fact, it
is important to know the degree of divergence in the correlation
functions.
First of all, since the propagator is defined on a compact manifold,
infrared divergencies are absent.
Choosing the Feynman gauge, $\lambda=1$, one picks up in eqs.
\gzw-\gzwb\ the components $G_{\bar z w}(z,w)$ and $G_{z\bar
w}(z,w)$. In analogy with the flat case,  we expect that the
propagator in the Feynman gauge
has a logarithmic singularity at short distances.
This implies that
the derivatives of the propagator should have a simple pole when
$z\rightarrow w$.
Indeed, deriving eq. \gzbw\ in $z$ and using the property \kone\ of
the scalar Green function $K(z,w)$, one obtains:
\eqn\singfey{
\partial_zG_{\bar z w}(z,w)=-{1\over 2}g_{z\bar z}\partial_wK(z,w)+
{1\over 2A}g_{z\bar z}\int_Md^2tg_{t\bar t}\partial_wK(w,t)}
Clearly, the right hand side has a simple pole, since
$\partial_wK(z,w)\sim {1\over z-w}$. No other divergencies are present
in eq. \singfey\ because the second term in the right hand side
vanishes
due to eq. \kthree.
An analogous result holds in the case of
$G_{z\bar w}(z,w)$.\smallskip
Now we consider the components $G_{zw}(z,w)$ and $G_{\bar z\bar w}(z,w)$.
They are picked up choosing the gauge $\lambda=-1$.
The possible divergencies may arise only in the limit $z\rightarrow w$.
However, a simple look at eqs. \flatgzw\ and \flatgzbwb\ shows that
there are no poles in this limit at least in the flat case.
As a matter of fact, the expression of the biharmonic Green function
$1\over \triangle^2$
at short distances is given by:
\eqn\bgflat{G_{flat}(z,w)\sim {1\over 2}|z-w|^2{\rm log}|z-w|+\ldots}
{}From the above formula, it is clear that
$\partial_z\partial_wG_{flat}(z,w)$ and $\partial_{\bar z}
\partial_{\bar w}G_{flat}(z,w)$ do not have any divergence
when $z\rightarrow w$.
The finiteness of these components is also clear from the expression
of the propagator \flatprop\ in the Fourier space.
This is just an accident, caused by the fact that the
logarithmic divergence of $G_{flat}(z,w)$
in $z=w$ is hidden by the factor $|z-w|^2$.
Indeed, $G_{zw}(z,w)$
and $G_{\bar z\bar w}(z,w)$ remain distributions and the
singularities emerge after exploiting the equations of motion
\propone-\proptwo.
Since the short distance behavior of the correlation
functions should not depend on the topology, we expect that the
finiteness of the components holds not only in the flat case, but also on a
Riemann surface of any genus.
To prove this statement, we rewrite the integral in \gzw\ as follows:
$$
\int_Md^2tg_{t\bar t}
\partial_zK(z,w)\partial_wK(z,w)=$$
\eqn\intsplit{
\int_Dd^2tg_{t\bar t}
\partial_zK(z,w)\partial_wK(z,w)+\int_{M-D}
d^2tg_{t\bar t}\partial_zK(z,w)\partial_wK(z,w)}
where $D$ is a small disk of radius $\epsilon$ cut in
the Riemann surface. $D$ contains both the
points $z$ and $w$, which are supposed to be very close.
The second integral in the right hand side of eq. \intsplit\ is harmless and
the potential singularities are present only in the first integral over the
disk
$D$, where $K(z,w)\sim {\rm log}|z-w|$.
As a consequence, taking a locally flat metric on $D$,
the leading divergent term is:
\eqn\mdivapprox{\int_Dd^2tg_{t\bar t}\partial_zK(z,w)\partial_wK(z,w)\sim
\int_Dd^2t{1\over (t-z)^2}+O(z-w)}
Using a system of polar coordinates $r$ and $\theta$
centered at the point $z$,
the above integral becomes (see also ref.
\ref\gs{I. M. Gelfand and G. E. Shilov, Generalized Functions I, Academic
Press,
New York and London 1964.}, pag.375):
\eqn\mdivzero{
\int_D{dtd\bar t\over (t-z)^2}=-2i\int_0^\epsilon r dr\int_0^{2\pi}e^{-2i
\theta} d\theta=0}
Therefore, inserting eq. \mdivzero\ in eq. \mdivapprox\ and substituting
again the
latter into eq. \intsplit, it turns out that
$G_{zw}(z,w)$ remains finite in the limit $z\rightarrow w$.
This result is independent of the fact that we have used the particular
topology of a disk. The choice of another simply connected manifold with
boundary amounts in fact only to a conformal transformation, which is
irrelevant
in eq. \intsplit, because it is written in a covariant way.
An analogous proof can be performed also in the case of
$G_{\bar z\bar w}(z,w)$.\smallskip
The finite parts of $G_{zw}(z,w)$ and
$G_{\bar z\bar w}(z,w)$ may also play a role in perturbation theory.
In order to compute them, one has to evaluate the following
integrals:
\eqn\gzz{G_{zz}(z,\bar z)=\int_Md^2tg_{t\bar
t}\left[\partial_zK(z,t)\right]^2}
\eqn\gzbzb{G_{\bar z\bar z}(z,\bar z)=\int_Md^2tg_{t\bar
t}\left[\partial_{\bar z}K(z,t)\right]^2}
$G_{zz}(z,\bar z)$ and $G_{\bar z\bar z}(z,\bar z)$ should be
singlevalued tensors on a
Riemann surface without singularities.
The strategy exploited in order to solve these integrals is to
rewrite the integrand in another form, which reproduces the poles
of $\partial_zK(z,t)^2$ at  $z=t$ but is linear in $K(z,w)$.
For instance, we start with the sphere of genus zero $S^2$.
Choosing the metric $g_{z\bar z}dzd\bar z={dzd\bar z\over (1+z\bar
z)^2}$ the scalar Green function $K(z,w)$ becomes:
\eqn\kzwsphere{K(z,w)={\rm log}\left[{|z-w|^2\over(1+z\bar z)(1+w\bar
w)}\right]}
and one can apply this formula in eqs. \gzw-\gzbwb\ in
order to obtain the explicit form of the propagator. Moreover, from
eq. \kzwsphere\ we infer the following nice identity:
\eqn\spheretras{\left[\partial_zK(z,t)\right]^2=-\nabla_z\partial_zK(z,t)}
where $\nabla_z=\partial_z+g_{z\bar z}\partial_zg^{z\bar z}$ is the
covariant derivative acting on the $(1,0)-$forms.
Substituting eq. \spheretras\ in eq. \gzz\ and exploiting the
properties of the scalar Green function $K(z,w)$, in particular eq.
\kthree, one obtains:
\eqn\dvipropst{G_{zz}(z,\bar z)=\int_{S^2}d^2tg_{t\bar
t}\nabla_z\partial_zK(z,t)=0}
An analogous result holds for $G_{\bar z\bar z}(z,\bar z)$.
\smallskip
On the torus the computation of $G_{zz}(z,\bar z)$ and
$G_{\bar z\bar z}(z,\bar z)$ is very simple due to the translational
invariance of the
scalar Green function $K(z,w)\equiv K(z-w)$.
As a matter of fact,
choosing a flat metric $g_{t\bar t}=1$ in eqs. \gzz\ and \gzbzb,
one can perform the substitution $t'=z-t$ and set $\partial_zK(z-t)=
-\partial_tK(z-t)$.
The upshot is that $G_{zz}(z,\bar z)$ and
$G_{\bar z\bar z}(z,\bar z)$ are constants given by:
$$G_{zz}(z,\bar
z)=\int_Md^2t'\left[\partial_{t'}K(t')\right]^2\qquad\qquad
G_{\bar z\bar z}(z,\bar
z)=\int_Md^2t'\left[\partial_{\bar t'}K(t')\right]^2$$
On the Riemann surfaces of genus $g>1$, however, there is no translational
invariance, so that the tensors
$G_{zz}(z,\bar z)$ and
$G_{\bar z\bar z}(z,\bar z)$ receive a dependency on $z$.
Their expression will be explicitly computed in appendix A.
\vfill\eject
\newsec{5 Yang$-$Mills Field Theories}

In the previous sections the propagators of Yang$-$Mills field
theories quantized in the covariant gauge have been explicitly
computed on any Riemann surface of genus $h$.
Adding also the color indices, which
play however an irrelevant role in the free equations of motion
\propone-\propfour,
the components of the propagator read:
\eqn\gzwab{G_{zw}^{ab}(z,w)=-\delta^{ab}{\lambda-1\over 4}\int_Md^2tg_{t\bar
t}\partial_zK(z,t)\partial_wK(w,t)}
\eqn\gzbwab{G_{\bar zw}^{ab}
(z,w)=-\delta^{ab}{\lambda+1\over 4}\int_Md^2tg_{t\bar
t}\partial_{\bar z}K(z,t)\partial_wK(w,t)}
\eqn\gzbwbab{G_{\bar z\bar w}^{ab}
(z,w)=-\delta^{ab}{\lambda-1\over 4}\int_Md^2tg_{t\bar
t}\partial_{\bar z}K(z,t)\partial_{\bar w}K(w,t)}
\eqn\gzwbab{G_{z\bar w}^{ab}(z,w)=
-\delta^{ab}{\lambda+1\over 4}\int_Md^2tg_{t\bar
t}\partial_zK(z,t)\partial_{\bar w}K(w,t)}
Analogously we have for the ghost fields:
\eqn\ggh{G_{gh}^{ab}(z,w)=\delta^{ab}K(z,w)}
where $K(x,y)$ is the scalar Green function \kone.\smallskip
Unfortunately, the knowledge of the propagators alone is not sufficient
on a Riemann surface in order to compute all the other correlation functions
perturbatively.
The second necessary ingredient is provided by the flat connections
\connections, \bt, \hitchin\ and \sonnen.
In complex coordinates, they are given by the
independent solutions of the equation:
\eqn\flatconn{\partial_zA_{\bar z}-\partial_{\bar z}A_z+i\kappa
[A_z,A_{\bar z}]=0}
which can be constructed as follows
(see also the Appendix of ref. \sonnen). We consider the
$2h(N^2-1)$ independent gauge fields $A_z^{(i),\bar a}(z)$ and
$A_{\bar z}^{(i),\bar a}(\bar z)\equiv (A_z^{(i),\bar a}(z))^*$ defined by:
\eqn\azflat{A_z^{(i),\bar a}(z)=\omega_i(z)\delta^{\bar a b}
T^b\qquad\qquad\qquad
A_{\bar z}^{(i),\bar a}(\bar z)=\bar\omega_i(\bar z)\delta^{\bar a b}(T^b)^*}
where $i=1,\ldots,h$ and $\bar a=1,\ldots,(N^2-1)$.
In the usual representation of the connections as $su(N)$ valued vector
fields $A_\alpha=A_\alpha^bT^b$, we have that
$(A_z^{(i),\bar a}(z))^b=\omega_i(z)\delta^{\bar a b}$ and
$(A_{\bar z}^{(i),\bar a}(\bar z))^b=-\bar\omega_i(\bar z)\delta^{\bar a
b}$.
Thus $\bar a$ labels the possible independent solutions of eq.
\flatconn\ and simultaneously is also a color index.\smallskip
We recall that the $T^a$ are in the adjoint representation, so that we can
use here the standard form of the $SU(N)$ generators
$(T^a)_{ik}=if^{aik}$. In this way the $f^{abc}$ turn out to be real
structure constants from the commutation relations
$[T^a,T^b]=if^{abc}T^c$ and
the elements of the totally antisymmetric matrices $T^a$ are purely
imaginary, i.e. $(T^a)^\dagger=T^a$ and $(T^a)^*=-T^a$.
It is now clear that the commutator $[A_z^{(i),\bar a}(z),A_{\bar z}^{(i),
\bar a}(\bar
z)]$ vanishes, because
$[T^a,(T^a)^*]=-[T^a,T^a]=0$
and therefore
$$[A_z^{(i),\bar a}(z),A_{\bar z}^{(i),\bar a}(\bar z)]
=-[T^{\bar a},T^{\bar a}]\enskip\omega_i(z)\bar\omega_i(\bar z)=0$$
Moreover, since the $\omega_i(z)$ and $\bar\omega_i(\bar z)$ are
abelian differentials, the following identity is valid:
\eqn\linzero{\partial_z
A_{\bar z}^{(i),\bar a}-\partial_{\bar z}A_z^{(i),\bar a}=0}
Hence, we have shown that the differentials $A_z^{(i),\bar a}(z)$ and
$A_{\bar z}^{(i),\bar a}(\bar z)$ satisfy eq.
\flatconn. Exploiting the freedom of performing gauge
transformations of the kind \gaugetransf, the
most general expression of these flat connections will be:
\eqn\azflatfin{\tilde A_z^{(i),\bar a}(z,\bar z)=U^{-1}[A^{(i),\bar
a}_z(z)-i\kappa^{-1}\partial_z]U}
\eqn\azbflatfin{\tilde A_{\bar z}^{(i),\bar a}(z,\bar z)=U^{-1}[A^{(i),\bar
a}_{\bar z}(\bar z)
-i\kappa^{-1}\partial_{\bar z}]U}
We notice at this point that the
$2h(N^2-1)$ special flat connections given above
are apparently independent, but some degrees of freedom
can still be eliminated by means of the gauge
transformations \azflatfin\ and \azbflatfin.
The dimension of the moduli space of flat connections $M_F(M,SU(N)))$
is indeed $(2h-2)(N^2-1)$.
A proof of this fact, extended also to the more general
self-dual connections, is
in ref. \hitchin.
In our particular case the dimensionality of $M_F(M,SU(N)))$
does not play an important role, since we are only interested in the
perturbative expansion of the Yang$-$Mills amplitudes near a classical
configuration $A^{cl}_\alpha(z,\bar z)$ satisfying eq. \flatconn.
Clearly, $A_\alpha^{cl}(z,\bar z)$ can be always written as a linear
combination of the basis \azflat. Accordingly to our strategy, we
expand the gauge fields as follows:
\eqn\clpqu{A_\alpha(z,\bar z)=A_\alpha^{cl}(z,\bar
z)+A_\alpha^q(z,\bar z)}
where $A^q_\alpha(z,\bar z)$ describes a quantum fluctuation around
$A^{cl}_\alpha$.\smallskip
To quantize the theory,
it is now possible to
proceed as in the previous sections, imposing the covariant
gauge \confgauge\ only on the quantum perturbation $A^q_\alpha$.
As an upshot, the ghost action and the gauge fixing term \brsvar\ do
not contain $A_\alpha^{cl}$ and
the generating functional is the same of eq.
\genfunct:
\eqn\genfunc{Z[J]=\int DA_\mu^q D\bar cDc \enskip
e^{\left\{-{\rm Tr}\int_Md^2x\sqrt{g}
\left[{1\over 4}F^2(A)+{1\over
2\lambda}f^2(A^q)+\partial_\mu \bar cD^\mu(A^q)c+J^\mu A_\mu^q\right]\right\}}}
apart from the replacement:
$F_{\mu\nu}(A)\equiv F_{\mu\nu}(A^{cl}+A^q)$.
\smallskip
In this way, however, the invariance of the amplitudes
under gauge transformations of
$A^{cl}_\mu$ is lost.
To remedy, one can
apply the techniques of refs. \background,
choosing the background gauge fixing:
$$f'(A^q,A^{cl})={1\over \sqrt{g}}
\partial_\mu(
\sqrt{g} A^\mu)+\kappa f^{abc}\tilde A^{(i)b}_\mu A^{\mu c}=0$$
Since this gauge fixing is not affecting the free part of the action
(when $\kappa=0$ it coincides with the covariant gauge \covgauge),
the free propagators of the theory can be computed as before and are
given again by eqs. \gzwab-\ggh.
\newsec{6 Conclusions}

The main result of this paper is the calculation of the relevant
propagators entering in Yang$-$Mills field theories defined
on a Riemann surface of any genus.
In particular, we have shown that
the requirements a)-c) of Section 3 determine the
propagator of the gauge fields uniquely.
As a proof of the physicality of our propagators, the Slavnov-Taylor
identity \fbrseq\ has been verified.
We would like to notice
that on a Riemann surface only $exact$ and $coexact$
forms propagate, while the notion of particles is lost.
{}From our investigations two unexpected
results emerge.
First of all, in complex coordinates not only
the Feynman gauge, but also the gauge $\lambda=-1$
is very suitable for calculations.
Moreover, we have used here a covariant gauge fixing, but the analysis
of Section 2 indicates that there is also the interesting
possibility of
quantizing the Yang$-$Mills theories on a compact two
dimensional manifold in a noncovariant gauge. As a matter of
fact, starting from a metric which is not conformally flat, we are
still allowed to impose the gauge fixing \confgauge. The reason is that
eq. \confgauge\ is compatible with the holomorphic transition functions
on the Riemann surface and can be globally extended over the entire manifold.
Involving only the component $g_{z\bar z}$ of the metric,
this gauge fixing destroys
the covariance of the pure Yang$-$Mills functional under global
diffeomorphisms. We remark that this procedure of choosing gauge has no
analogous in the flat space. In particular, more classical noncovariant gauges,
like for instance the axial gauge \ref\axial{R. L. Arnowitt and S. I. Flicker,
{\it Phys. Rev.} {\bf 127} (1962), 1821.}, the Coulomb gauge or the light
cone gauge \ref\thooft{G. 't Hooft, {\it Nucl. Phys.} {\bf B75}
(1974), 461.},
are not suitable in our case because they cannot
be globally imposed on $M$.
\smallskip
With the expressions given here for the propagators
it is possible to start the computation of the other
correlation functions and of their radiative corrections.
The contributions coming from the flat connections can be evaluated by means
of the explicit formulas \azflat. Many simplifications
are expected to occur in the amplitudes because, due to requirement a), it is
easy to see that the gauge propagator \gzwab-\gzwbab\ is orthogonal
with respect to the flat connections. Moreover, most of the
physically relevant two dimensional models,
like Quantum chromodynamics,
are superrenormalizable. For instance, in the pure Yang$-$Mills
case, there is only one logarithmically
divergent Feynman diagram, corresponding to the one-loop
correction of the two point function.
Using the fact that on a compact manifold all the possible singularities
are ultraviolet, so that they occur at short distances where the topology
does not play any particular role, it should not be difficult to
subtract suitable counterterms in the Lagrangian in order to achieve a finite
theory. This would be an important result, proving the
renormalizability of gauge field theories on every closed
and orientable Riemann surface in an explicit and direct way.
However, the computability of the divergent Feynman integrals should still
be improved. This is not a simple problem. Even in the case of string theory,
explicit calculations have been performed only representing the Riemann
surface as an algebraic curve, i.e. as an $n$ sheeted covering of the
complex plane
\ferstr, \ref\knizhnik{
V. G. Knizhnik, {\it Sov. Phys. Usp.} {\bf 32} (11) (1989), 945;
M. A. Bershadsky and A. O. Radul, {\it Int. Jour. Mod. Phys.} {\bf A2} (1987),
165; J. Sobczyk, {\it Mod. Phys. Lett.} {\bf A6} (1991), 1103.},
\ref\algcur{
D. Lebedev and A. Morozov, {\it Nucl. Phys.} {\bf B302} (1988), 63;
A. A. Belavin, V. G. Knizhnik, A. Yu. Morozov and A. M. Perelomov,
{\it JETP Lett.} {\bf 43} (1986), 411;
D. J. Gross and P. F. Mende, {\it Nucl. Phys.} {\bf B303} (1988), 407;
R. Iengo, {\it Nucl. Phys.} {\bf B15} (Proc. Suppl.) (1990), 67;
E. Gava, R. Iengo and G. Sotkov, {\it Phys. Lett.} {\bf 207B} (1988), 283;
D. Montano, {\it Nucl. Phys.} {\bf B297} (1988), 125.},
\ref\ferstrtwo{F. Ferrari and J. Sobczyk, Operator Formalism on General
Algebraic Curves, Preprint U.T.F. 333, IFT UWr 879/94;
F. Ferrari, {\it Int. Jour. Mod. Phys.} {\bf A7} (1992), 5131; J. Sobczyk and
W. Urbanik, {\it Lett. Math. Phys.} {\bf 21} (1991), 1; J. Sobczyk,
{\it Mod. Phys. Lett.} {\bf A6} (1991), 1103.}.
An important step in this direction would be the construction of the biharmonic
Green function on any algebraic curve, which is currently under investigation
\ref\progress{F. Ferrari and J. Sobczyk, work in progress.}.
Recently the Schwinger model quantized in the Lorentz gauge
has been successfully
solved on any Riemann surface within our explicit
formalism, computing the correlation
functions of the fermionic currents in a nonperturbative way \fersch, \ferabe.
We hope therefore that, with the material presented here,
it will be possible
to extend these results also to the Yang$-$Mills field theories.
\appendix{Appendix A}

In this appendix the explicit form of the tensor $G_{zz}(z,\bar z)$ of eq.
\gzz\ will be computed.
We start from the following formula
\ref\ks{S. M. Kuzenko and O. A. Solov'ev, JETP 51 (1990), 265.},
which generalizes eq. \spheretras\ to any Riemann surface:
$$\left[\partial_zK(z,t)\right]^2=\partial_z^2K(z,t)+2\partial_z
K(z,t)\int_Md^2yg_{y\bar y}\partial_zK(z,y)\tilde
R_g(y)-$$
$$2\int_Md^2y\partial_zK(z,y)\partial_yK(t,y)P_{z\bar
y}+\Psi_{zz}(z,t)+$$
\eqn\kktok{{1\over A}\int_Md^2vg_{v\bar v}\partial_zK^{(+)v}_z(z,v)
\int_Md^2yg_{y\bar y}\partial_vK(v,y)\tilde R(y)}
where
$$\tilde R_g(z)\equiv R_g(z)+{1\over 2}
g^{z\bar z}\sum\limits_{i,j=1}^h\omega_i(z)
\left[{\rm Im}\enskip\Omega\right]_{ij}^{-1}\bar\omega_j(\bar z)$$
and
$$P_{z\bar y}\equiv{1\over 2}\sum\limits_{ij=1}^h\omega_i(z)\left[{\rm
Im}\enskip\Omega\right]_{ij}^{-1}\bar\omega_j(\bar y)$$
Moreover the Green function $G^{(+)v}_z(z,v)$ satisfies the equation
(see also \ref\dhp{E. D'Hoker and D. H. Phong, {\it Rev. Mod. Phys.} {\bf
60} (1988), 917.}):
$$\triangle_1^{(+)}G_z^{(+)v}(z,w)=\delta^{(2)}(z,v)$$
and finally
$\Psi_{zz}(z,t)$ is a linear combination of the
$3h-3$ holomorphic quadratic
differentials with coefficients depending on  $t$.
We notice also that our formula is slightly different to that of \ks\
in order to take into account of the different normalization of the
scalar Green function $K(z,w)$ given in eqs. \kone\ and \ktwo.
Substituting eq. \kktok\ in eq. \gzz, and exploiting the property
\kthree, one easily proves that
\eqn\gzzexpl{G_{zz}(z,\bar z)=
\Psi_{zz}(z)
+\int_Md^2vg_{v\bar v}\partial_zK^{(+)v}_z(z,v)
\int_Md^2yg_{y\bar y}\partial_vK(v,y)\tilde R(y)}
An analogous formula can be found for $G_{\bar z\bar z}(z,\bar z)$.
As anticipated in section 3, the lack of translational invariance
yields a form of $G_{zz}(z,\bar z)$ and $G_{\bar z\bar z}(z,\bar z)$
which is dependent on the space-time variables.

\immediate\closeout\rfile\writestoppt
\section {References}

\input refs.tmp\vfill\eject
\end